\documentclass[10pt]{article}

\usepackage{graphics}
\input{epsf}
\usepackage{euscript,amssymb}

\font\msbm=msbm10

\catcode`\@=11
\def\lesssim{\mathrel{\mathpalette\vereq<}}
\def\vereq#1#2{\lower3pt\vbox{\baselineskip1.5pt \lineskip1.5pt
\ialign{$\m@th#1\hfill##\hfil$\crcr#2\crcr\sim\crcr}}}

\def\Let@{\relax\iffalse{\fi\let\\=\cr\iffalse}\fi}
\def\vspace@{\def\vspace##1{\crcr\noalign{\vskip##1\relax}}}
\def\multilimits@{\bgroup\vspace@\Let@
 \baselineskip\fontdimen10 \scriptfont\tw@
 \advance\baselineskip\fontdimen12 \scriptfont\tw@
 \lineskip\thr@@\fontdimen8 \scriptfont\thr@@
 \lineskiplimit\lineskip
 \vbox\bgroup\ialign\bgroup\hfil$\m@th\scriptstyle{##}$\hfil\crcr}
\def\Sb{_\multilimits@}
\def\endSb{\crcr\egroup\egroup\egroup}
\def\Sp{^\multilimits@}

\newcommand{\be}[1]{\begin{equation}\label{#1}}
\newcommand{\ee}{\end{equation}}
\newcommand{\ba}[1]{\begin{eqnarray}\label{#1}}
\newcommand{\ea}{\end{eqnarray}}
\newcommand{\rf}[1]{(\ref{#1})}
\newcommand{\nn}{\nonumber}

\newcommand{\sn}{\mbox{\rm sn}}

\newcommand{\cd}{\mbox{\rm cd}}
\newcommand{\ds}{\mbox{\rm ds}}
\newcommand{\const}{\mbox{\rm const}}

\newcommand{\arctanh}{\mbox{\rm arctanh}}

\begin{document}

\author{Mariam Bouhmadi--L\'
opez\footnote{e-mail: mbouhmadi@imaff.cfmac.csic.es}, Pedro F.
Gonz\'{a}lez--D{\'\i}az\footnote{e-mail:
p.gonzalezdiaz@imaff.cfmac.csic.es} \\ and Alexander
Zhuk\footnote{e-mail: ai$\_$zhuk@imaff.cfmac.csic.es
\newline
on leave from: Department of Physics, University of Odessa, 2
Dvoryanskaya St., Odessa 65100, Ukraine} \\ \\ Instituto de
Matem\'{a}ticas y F\'{\i}sica Fundamental, \\ Consejo Superior de
Investigaciones Cient\'{\i}ficas,
\\ C/ Serrano 121, 28006 Madrid, Spain }

\title{On new gravitational instantons describing creation of brane-worlds}

\date{2002}
\maketitle

\abstract{By considering 5--dimensional cosmological models with a
bulk filled with a pressureless scalar field; equivalently dust
matter, and a negative cosmological constant, we have found a
regular instantonic solution which is free from any singularity at
the origin of the extra--coordinate. This instanton describes
5--dimensional asymptotically anti de Sitter wormhole, when the
bulk has a topology $\msbm R \times S^4 $. Compactified
brane-world instantons which are built up from such instantonic
solution describe either a single brane or a string of branes.
Their analytical continuation to the pseudo--Riemannian metric can
give rise to either 4-dimensional inflating branes or solutions
with the same dynamical behaviour for extra--dimension and branes,
in addition to multitemporal solutions. Dust brane-world models
with arbitrary dimensions ($D\ge 5$) as well as other spatial
topologies are also briefly discussed. }

\bigskip

\hspace*{0.950cm} PACS number(s): 04.50.+h, 98.80.Hw



\section{Introduction}
\setcounter{equation}{0}

\bigskip

Although higher dimensional cosmological models can be traced back
to the first years of gravitation theory, the idea has received a renewed
great attention in the last few years, due to the publication of pioneering
papers on brane and extra-dimension models which shed light for the solution
of fundamental physical problems including the extra-dimension
compactification and the hierarchy problem. Extra-dimensions can be
compactified, as in the
standard Kaluza-Klein theory \cite{KK}, or not, as was firstly suggested by
Akama \cite{Akama},
Rubakov and Shaposhnikov \cite{Ru-Sh} and others
\cite{Visser}.
While Akama \cite{Akama} considered the universe as a four vortex-like object
embedded in a 6--dimensional flat space-time, Rubakov and Shaposhnikov
proposed a model in (1+N)--dimensional Minkowski space-time ($N\geq 4$) where
particles were confined in a potential well, flat on the usual three spatial
dimensions and narrow along the extra-dimensions. Gravity was later on
included in 5--dimensional manifolds \cite{Visser} in order to trap particles
near a 4--dimensional Lorentzian submanifolds.

More recently, a model was proposed \cite{ADD} on
(4+N)--dimensions with $N\geq2$, where the extra-dimensions are
compact and their size, $R$, is deduced by imposing that the usual
Planck scale, $M_{Pl}$, is no longer a fundamental scale and
Planck scale in (4+N)--dimensions, $M_{Pl_{4+N}}$, is of the order
of the weak scale $M_{EW}$, $M_{Pl}^2=M_{Pl_{4+N}}^{2+N}R^N$. This
model can solve the hierarchy problem. In this framework the
gravitons can propagate in the extra--dimensions while the
standard model fields are confined to a 4--dimensional
submanifolds of thickness $M_{EW}^{-1}$ in the extra--dimensions.

Randall and Sundrum suggested a new approach to solve the
hierarchy problem \cite{RS1} by including just one extra compact
dimension. In their first model \cite{RS1}, inspired by string
theory \cite{Witten}, they considered 5--dimensional anti de
Sitter (AdS) bulk with two branes with opposite tension; our
universe is then placed on the brane with negative tension where
standard model particles are localized. In a second model
\cite{RS2}, these authors placed the universe in the brane with
positive tension, in a non-compact background. In this framework,
it is possible to reproduce 4--dimensional general relativity even
if the extra--dimension is non-compact \cite{RS2}, due to the
existence of a massless gravitational bound state of Kaluza-Klein
(KK) modes which is the graviton of 4--dimensional world. While
for the non-compact case, KK spectrum is continuous without a gap,
for the compact case the KK excitations are quantized.

Branes in the Randall-Sundrum models are 4--dimensional flat
space-time and consequently, at least in principle, they can not
describe any inflationary universe. Nevertheless, as it was
pointing out by Garriga and Sasaki (GS) \cite{Gariga-Sasaki} it is
still possible to construct an inflating brane, whose geometry
corresponds to a 4--dimensional de Sitter space, surrounded by a
5-dimensional AdS. The Euclidean version of this solution can be
used for the description of the creation of the universe from
nothing. The GS model has a single normalized gravitational bound
state which corresponds to the massless graviton and separates by
a gap from the massive KK modes.

Multi-brane-world models also exhibit 4--dimensional gravity
localization at the branes. In particular, a compact brane-world
model consisting of three flat branes, two with positive tensions
and an intermediate one with negative tension, embedded in
5-dimensional AdS space-time has been also considered
\cite{Santiago}. In this case, the universe is placed in one of the
positive tension branes. In addition, intersecting brane
configurations \cite{ADDK}, crystal brane-world \cite{Kaloper} and
brane periodic configuration \cite{Nam} have been investigated as
well. The massless gravitational bound state exists in all these
models.

However, most of the above mentioned brane-world models have the
shortcoming of either not including creation of inflating branes
or having a singular character for brane instantons. The main aim
of the present paper is to propose a set of singularity free
5--dimensional models which are able to produce inflating
brane-worlds. This can be achieved by considering matter in the
bulk which we model by a scalar field. As we shall see, such
models should also induce localization of 4--dimensional gravity
at the branes.

The paper is organized as follows. In the next section we derive a
master equation \rf{2.9} which describes the behaviour of the
scale factor of D--dimensional instantonic (after an analytic
continuation to the Euclidean region) solutions for a bulk filled
with a scalar field, which is subjected to a perfect fluid state
equation, and a cosmological constant, $\Lambda_{D}$. For $D=5$,
$\Lambda_{D}<0$ and spherical 4-dimensional sections, we obtain
asymptotically AdS wormhole when the bulk is filled with dust.
Using this wormhole solution, in section 3, we construct compact
and non-compact brane-world instantons. In section 4, we describe
the brane-world birth from nothing (the Euclidean solution)
performing an analytical continuation of the instantonic solution.
In section 5, we analyze the behaviour of the massless
gravitational KK mode for the models constructed in the previous
section. Finally, in section 6, we summarize our results and
discuss other dust brane-world models with arbitrary dimensions
($D\ge 5$) as well as other spatial topologies.

\section{Multidimensional perfect fluid cosmology \label{setup}}
\setcounter{equation}{0}

\bigskip

Let us start our investigation of a multidimensional model with a
cosmological constant, $\Lambda_D$, and minimal scalar field,
$\varphi$, by writing down the action\footnote{Although the
constant minimal scalar field is really equivalent to a
cosmological term, it is convenient to single out an explicit
cosmological constant in the action. \label{1}}:
\ba{2.1} S &=& \frac{1}{2\kappa^2_D} \int d^{\,
D}X\sqrt{|g^{(D)}|}\left\{ R[g^{(D)}] - 2 \Lambda_D \right\}\\ &+&
\int d^{\, D}X\sqrt{|g^{(D)}|}\left\{ -\frac12 g^{(D)MN}
\partial_{M} \varphi
\partial_{N} \varphi - V(\varphi ) \right\} + S_{YGH} \, , \nn  \ea
where $\kappa^2_D$ is D--dimensional gravitational constant and
$S_{YGH}$ the standard York-Gibbons-Hawking boundary term
\cite{YGH}. The $(D=1+d)$--dimensional metric $g^{(D)}$ is taken
to have the form
\be{2.2} g^{(D)} = g^{(D)}_{MN}dX^M\otimes dX^N = - e^{2\gamma
(\tau )} d\tau \otimes d\tau + e^{2\beta (\tau )} g^{(d)}_{\mu
\nu} dx^{\mu} \otimes dx^{\nu}\ , \ee
with $g^{(d)}$ the metric of d-dimensional Einstein space:
$R[g^{(d)}] = \lambda d \equiv R_d$, and in the case of constant
curvature space parameter $\lambda$ is normalized as $\lambda =
k(d-1)$, with $k=\pm 1,0$.

Consistent with the metric ansatz
\rf{2.2}, we suppose that the scalar field is also homogeneous:
$\varphi = \varphi (\tau )$. For such a scalar field energy
density and pressure are defined as:
\be{2.3} \vphantom{\int} \begin{array}{rcl} T^0_0 &=& -\frac12
e^{-2\gamma} \dot \varphi^2 -V(\varphi )\, \equiv \; -\rho\, ,
\phantom{\frac{R^l_i}{d^l_i}} \\ T^{\mu}_{\mu} &=& \quad \frac12
e^{-2\gamma} \dot \varphi^2 -V(\varphi )\, \equiv \; P\, , \qquad
\mu = 1, \ldots ,d \, ,\phantom{\frac{R^l_i}{d^l_i}} \\
\end{array}
\ee
and supposed to satisfy the state equation:
\be{2.4} P = (\alpha - 1)\rho\, . \ee
The conservation equation $T^M_{N;M} = 0$ implies $\rho=A
a^{-4\alpha}$ where $A$ is an arbitrary constant and $a(\tau ) :=
\exp [\beta (\tau )]$ is a scale factor. This results in a inverse
power-law for the scalar field potencial $V$ in term of the scale
factor $a= \exp[\beta(\tau)]$; i.e.
$V=(1-\frac{\alpha}{2})A\,a^{-d\alpha}$, whenever $\alpha>0$.
Obviously, $a^d$ gives the volume of the Universe at any
hypersurface $\tau = \const$ up to a spatial prefactor $V_d = \int
d^dx \sqrt{|g^{(d)}|}$.

{}From the energodominant condition it is usually supposed that
$\rho \ge 0$ and $-\rho \le P \le \rho$, where the latter
condition (which results in $0 \le \alpha \le 2$) means that the
speed of sound is less than the speed of light. For the aim of
generality we will not restrict ourselves to satisfy these
conditions because at least the first of them can be violated on
the case of a scalar field . For example, $\alpha = 0$ with $\rho
< 0$ describes the case of a negative cosmological constant.
Depending on the values acquired by the parameter $\alpha$, the
scalar field can mimic different kinds of matter. However, the
state equation Eq.~(\ref{2.4}) constrains strongly the form of the
potential. Below, we shall get, as an example, the potential
$V(\varphi)$ corresponding to a dust-like state equation ($\alpha
= 1$). Additionally, it can be shown \cite{Zhuk_QCG} that Einstein
equations for such scalar fields are equivalent to the Einstein
equations for a perfect fluid with action:
\be{2.6} S = \frac{V_d}{\kappa^2_D} \int d \tau \left[ \frac12
e^{-\gamma + \gamma_0} d(1-d) \dot \beta^2 - e^{\gamma - \gamma_0}
U \right] \equiv \frac{V_d}{\kappa^2_D} \int d \tau L \, , \ee
in which we performed integration over spatial coordinates $x$,
$\gamma_0 := d\beta$, overdot denotes differentiation with respect
to $\tau$ and we have introduced a "potential energy"
\be{2.7} U = e^{2 \gamma_0} \left( -\frac12 R_d e^{-2\beta} +
\Lambda_D + \kappa^2_D\, \rho \right)\, .\ee
The constraint equation corresponding to the Lagrangian in
Eq.\rf{2.6} results in the following master equation:
\be{2.8} \frac{\partial L}{\partial \gamma} = 0 \quad
\Longrightarrow \quad \frac12 e^{-\gamma + \gamma_0} d(1-d) \dot
\beta^2 + e^{\gamma - \gamma_0} U = 0 \, , \ee
which is exactly the 00--component of the Einstein equation. The
function $\gamma$ reflects the freedom for the choice of different
time gauges: $\gamma = 0$ is the proper time gauge, $\gamma =
\gamma_0$ is the harmonic time gauge \cite{IMZ} and $\gamma =
\beta$ is the conformal time gauge. Thus, in the proper time
gauge, Eq.\rf{2.8} reads
\be{2.9} \left(\frac{\dot a}{a}\right)^2 + \frac{R_d}{d(d-1)}
\frac{1}{a^2} + \frac{2}{d(1-d)} \Lambda_D + \frac{2}{d(1-d)}
\kappa^2_D A\, a^{-\alpha d} = 0 \, .\ee

We will concentrate mainly on 5--dimensional brane-world models
with AdS bulk: $D = 5, \, \Lambda_D = \Lambda_5 \equiv
-|\Lambda_5| <0$. The scalar curvature $R_d$ can be an arbitrary
constant but we shall consider the particular case of constant
curvature 4-D space $R_d = 12k$, restricting mainly to the
positive curvature with $k = +1$. For these parameters Eq.\rf{2.9}
becomes
\be{2.10} (\dot a)^2 + k + \Lambda a^2 - \bar A^2 \, a^{-4\alpha +
2} = 0\, , \ee
where for simplicity we have introduced the following notation:
$|\Lambda_5| /6 \equiv \Lambda$ and $(1/6) \kappa^2_5 A \equiv
\bar A^2$. Some of the particular values $\alpha$ are of special
interest because they correspond to important types of matter. For
example, in 5--D space-time\footnote{See e.g.
\cite{asrto-ph/9908047} and references therein for the discussion
of the equations of state for different kinds of cosmic defects in
usual 4--D universe. In these references a regime where the number
of defects per co-moving volume is a constant is considered.
\label{3}} $\alpha = 5/4$ describes radiation, $\alpha = 1$
corresponds to dust (0--D objects), $\alpha = 3/4$ represents
cosmic strings (1--D objects), $\alpha = 1/2$ describes domain
walls (2--D objects), $\alpha = 1/4$ corresponds to hyperdomain
walls (3--D objects) and $\alpha = 0$ represents vacuum (which can
be in some sense considered as a 4--D object).

Now, as we want to construct 5--D brane-world models, we perform
the Wick rotation to Euclidean "time" $r:\, \tau \to - ir$. Where
$r$ is to be considered as an extra spatial coordinate orthogonal
to 4--D branes, i.e., now hypersurfaces at $r = const$. Then, in
proper "time" gauge, metric \rf{2.2} is
\be{2.11} g^{(5)} = dr \otimes dr + a^2(r) g^{(4)} \, \ee
and the Euclidean version of Eq.\rf{2.10} reads
\be{2.12} (\dot a)^2 -k -\Lambda a^2 + \bar A^2 \, a^{-4\alpha +
2} = 0\, .\ee
Solutions of this equation describe 5--D instantons which can be
used to construct brane-world models. Obviously, the precise form
of the resulting instantons depends on the type of perfect fluid
we use, i.e., on the choice of the parameter $\alpha$. The vacuum
case $\alpha = 0$ (which here corresponds to a simple redefinition
of the cosmological constant: $\Lambda_D + \kappa^2_D A \to
\Lambda_D$) was considered in paper \cite{Gariga-Sasaki} for
positive curvature $k =+1$ with solution
\be{2.13} a(r) = l\, \sinh (r / l)\, , \ee
where $l := \sqrt{\Lambda^{-1}}$ is the AdS radius.

In the present paper we investigate the case of a pressureless
scalar field, i.e. with a dust-like state equation ($\alpha = 1$),
and $k =+1$. For this particular case the solution of Eq.\rf{2.12}
reads:
\be{2.14} a(r) = \frac{l}{\sqrt{2}} \left( \sqrt{b}\, \cosh
\frac{2r}{l} -1 \right)^{1/2}\, , \qquad -\infty < r < + \infty \,
, \ee
where $b := 1 + 4 \Lambda \bar A^{2}$. It can be easily seen that
this solution is reduced to \rf{2.13} in the limit: $\bar A^{2}
\to 0 \Longrightarrow b \to 1$. Metric \rf{2.11} with solution
\rf{2.14} is non-singular and describes a wormhole (the
integration constant being taken in such a way that $r=0$
corresponds to the wormhole throat). Having the solution
(\ref{2.14}), we can restore the dust-like scalar field potential
following the procedure described in \cite{Zhuk_QCG}. It yields
(in the Euclidean region):
\be{potential1} V(\varphi)= \frac{2A}{l^4}
\frac{(\sqrt{b}-1)\tanh^2\left(\bar \varphi\right)-\sqrt{b}-1}
{(\sqrt{b}-1)^2\tanh^2\left(\bar \varphi\right)+(\sqrt{b}+1)^2}\,
,\ee
where $\bar\varphi\equiv({l}/{2})\sqrt{{b-1}/{A}} \varphi$ and
runs from $-\arctanh \left [{\sqrt{b-1}}/({\sqrt{b}-{1}}) \right]$
to $\arctanh\left [{\sqrt{b-1}}/({\sqrt{b}-{1}})\right]$.

It is useful to present solution \rf{2.14} in the conformal "time"
gauge: $dr = \pm a d \eta$. It can be easily seen that
\be{2.15} a(\eta ) = l\, b^{1/4}\; {\ds\, (b^{1/4} |\eta |)}\; ,
\qquad 0\, \le\, b^{1/4} |\eta | \, \le \, 2K (m)\; ,\ee
satisfies the corresponding equation
\be{2.16} \left( \frac{d a}{d \eta} \right)^2 - a^2 -\Lambda a^4 +
\bar A^2 = 0 \, . \ee
Function $\ds\, u \equiv \ds \,(u|m)$ (with $m = (b^{1/2}
+1)/2b^{1/2}$ ) is a Jacobian elliptic function and $K (m)$ is the
complete elliptic integral of the first kind \cite{Abramowitz}.
For each of the sign of $\eta$, there is an one--one
correspondence between solution \rf{2.15} and wormhole \rf{2.14}.
Indeed, coordinates $r$ and $\eta$ are connected with each other
by :
\be{2.17} b^{1/4}\; |\eta | = F \left(\, \phi \, |\, m\, \right) =
u \quad \Longleftrightarrow \quad \sn\; b^{1/4} |\eta | =
\frac{1}{\cosh r/l} \; ,\ee
where $\phi := \arcsin (1/\cosh (r/l))$, $\sn\, u \equiv \sn \,
(u|m)$ is a Jacobian Elliptic function and $F \left(\, \phi \, |\,
m\, \right)$ is the incomplete elliptic integral of the first
kind. In the limit $\bar A^{2} \to 0 \Longrightarrow m\to 1\, $,
solution \rf{2.15} tends to $a(\eta ) \to l/ \sinh |\eta |$,
which, in fact, is solution \rf{2.13} expressed in the conformal
"time" gauge.

\section{Brane-world instantons\label{instanton}}
\setcounter{equation}{0}

\bigskip

In this section, we use solutions \rf{2.14} and \rf{2.15} to
construct brane-world instantons. This can be performed by
excising regions with $r > L$ for two identical wormholes
\rf{2.14} and gluing the remaining two copies along the two
4-spheres at $r=L$. The obtained instanton can be described by the
following piecewise continuous function:
\be{3.1} a (r) = \left\{\begin{array}{rcl} \frac{l}{\sqrt{2}}
\left[ \sqrt{b}\, \cosh \left(\frac{2r}{l}\right) -1
\right]^{1/2}\qquad &\, , &\quad -\infty < r \le L \\
\frac{l}{\sqrt{2}} \left[ \sqrt{b}\, \cosh
\left(\frac{2(2L-r)}{l}\right) -1 \right]^{1/2} &\, , & \qquad L
\le r < +\infty
\\
\end{array}\right. \ee
This function is continuous but not smooth at the gluing "point"
$r=L$. Due to the Lanczos-Israel junction condition, this results
in the appearance of a 4--D spherical brane at $r=L$, with a
tension:
\be{3.2} T(r=L)\, =\, \frac{1}{\kappa_5^2}\frac{3}{4} \widehat K
(L)\, = \, \frac{6}{\kappa_5^2 l}\; \frac{\sinh (L/l)\, \cosh
(L/l)}{\sinh^2 (L/l) + m_1}\, > \, 0\, , \ee
where $m_1 = 1-m\, ,\; \widehat K (L) \equiv K (L^+) - K (L^-)$,
and $K(L) = -4 a^{-1}(da/dr)_{r=L}$ is the trace of the extrinsic
curvature for the case of 5--D metric \rf{2.11}.

Instantonic solution \rf{3.1} is non-compact: the scale factor
$a(r)$ goes to $\infty$ when $r \to \pm \infty$, and cannot be
used for the description of the brane-world birth from "nothing"
(from the Euclidean region) because the probability of it in this
case is equal to zero. On the other hand, as we shall see below,
the bound state of the spin--2 gravitational perturbations
(corresponding to the Newtonian gravity on the brane) is
proportional to $a^{3/2}$ and is divergent when $a \to \infty$.
However, we can compactify this instanton identifying points
corresponding to the throats at $r=0 \leftrightarrow r=2L$. It
changes the topology along the extra dimension from $\mathbb{R}$
to $S^1$. Now, the range of the variation of $r$ is the interval
$[0,\, 2L]$. Since the geometry is smoothly glued at these points,
such procedure does not lead to the appearance of new
branes\footnote{Instanton \rf{3.1} can also be compactified by
additional cutting at distances $\triangle r=L_1 < L$ from the two
sides of the brane, gluing then along these two cuts. Geometry is
not smoothly matched at this surface which results in a new
negative tension brane. However, in our paper we will not consider
such a compact instanton because the first one, with the wormhole
throat identification, seems to us more elegant and it leaves a
possibility for 5--D baby-universe nucleation at the throat.
\label{4}}.

This procedure for the construction of the brane-world instanton
can be easily generalized to the case of an arbitrary number of
parallel branes by gluing one-brane manifolds at throats and
identifying the two final opposite throats. For example, in the
case of $n$ branes, located at the distances $r = L_i\, ,\; i= 1,
\ldots , n$ from throats, this instanton can be described by the
following piecewise continuous function:
\ba{3.3} a(r)& = & \sum_{i=1}^{n+1} a_i (r) \theta_i (r)\; ,
\qquad \qquad \qquad 0\, \le \, r \, \le \, 2 \sum_{i=1}^n L_i \;
,\\
 a_i(r)& = & \frac{l}{\sqrt{2}} \left[ \sqrt{b}\, \cosh
\left(\frac{2(2\sum_{k=1}^{i-1} L_k -r)}{l}\right) -1
\right]^{1/2}\, , \quad L_{i-1} +2\sum\limits_{k=1}^{i-2}L_k\; \le
\, r\, \le \; L_{i} +2\sum\limits_{k=1}^{i-1} L_k\; ,\nn \ea
where $L_0 \equiv L_{n+1} \equiv 0$ and
\be{3.4} \theta_i (r) =\tilde \eta (r-r_{i-1} ) -\tilde \eta
(r-r_i) =\left\{\begin{array}{rcl} &0&\, , \quad r < r_{i-1}\\
&1&\, , \quad r_{i-1} \le r < r_i \\ &0&\, , \quad r \ge r_i \\
\end{array}\right. \ee
are piecewise discontinuous functions, with $\tilde \eta (r -r_i)$
being the usual step function\footnote{ We put a tilde above it to
distinguish from the conformal time $\eta$. Obviously, $
\theta_i^{\, p} = \theta_i\; ,\; p>0\; ;\quad \theta_i \, \theta_j
= 0 \; ,\; i \ne j\; \Longrightarrow a^p = \sum_{i=1}^n a_i^p
\theta_i \; ,\; \forall p$ and $ \theta_i^{'} = \delta (r-r_{i-1})
- \delta (r-r_i)$. \label{5}} equal to zero for $r<r_i$ and
becoming unity for $r \ge r_i$. We have redefined the coordinate $r$ in
such a way to cover the range of variable for our solution by one
coordinate chart. In this case the i-th brane has coordinate $r=
r_i = L_i + 2\sum_{k=1}^{i-1}L_k\, , \; i = 1, \ldots ,n$ and the
i-th throat is located at $r= r_{(th)i} = 2\sum_{k=1}^{i-1}L_k\, ,
\; i = 1, \ldots ,n+1$. "Points" $r =r_0 \equiv r_{(th)1} =0$ and
$r = r_{n+1} \equiv r_{(th)n+1} = 2 \sum_{i=1}^n L_i$ are
identified with each other due to $S^1$-symmetry. Thus, metric
\rf{2.11} with scale factor \rf{3.3} describes a 5--D compact
instanton with $n$ parallel branes transversal to coordinate $r$
(see figure 1). Each of these branes has a tension given by
Eq.\rf{3.2} with the evident substitution $L \to L_i$ for the i-th
(i=1,\ldots ,n) brane.

The similar procedure for construction of brane-world instanton
can be performed in the conformal gauge. For example, one-brane
instanton can be obtained from the solution \rf{2.15} if we take
two wormholes with $\eta \lessgtr 0$ and $n=0$, cut them at $|\eta
| = \eta_0 < b^{-1/4} K(m)$, excise regions $|\eta | < \eta_0$ and
glue them along this cut. The obtained brane-world instanton is:
\be{3.5} a(\eta ) = l\, b^{1/4}\; {\ds\, \left[b^{1/4} (|\eta | +
\eta_0)\right] }\; ,\ee
where we redefined the coordinate $\eta $ in such a way ( $|\eta
|_{old} \to |\eta |_{new} + \eta_0$ ) to cover this instanton by
one coordinate chart. Here, $0 \le b^{1/4} |\eta | \le 2K(m) -
b^{1/4} \eta_0$ and $0 \le b^{1/4} |\eta | \le K(m) - b^{1/4}
\eta_0$ (with the identification $-b^{-1/4} K(m) +\eta_0
\leftrightarrow b^{-1/4} K(m) -\eta_0$) correspond respectively to
non-compact and compact instantons with brane at $\eta =0$ and
tension
\be{3.6} T(\eta =0)\, =\, \frac{6}{\kappa_5^2 l}\; \frac{\sqrt{1 -
\sn^2\, (b^{1/4}\eta_0})}{1 - m\; \sn^2\, (b^{1/4}\eta_0)}\,
> \, 0\, , \ee
which coincides with Eq.\rf{3.2} if $\eta = \eta_0$ and $r=L$ are
connected by formula \rf{2.17}.


\section{Brane-world birth from "nothing"\label{birth}}
\setcounter{equation}{0}

\bigskip

In this section we investigate the possibility for the creation of
 brane worlds from "nothing". To be more precise, we interpret
the brane-world instantons described in section 3 as a
semiclassical paths for the quantum tunneling from the Euclidean
region ("nothing").

Euclidean metric \rf{2.11} was obtained from the Lorentzian one by
the Wick rotation $\tau \to -ir$. For the compact solutions of
section 3 this Euclidean metric describes the compact 5--D
brane-world instantons. It is clear that back Wick rotation $r \to
i\tau$ will destroy these branes leaving, instead of the brane, a
solution corresponding to 5--D baby universe branching off from
the wormhole throats at $r=0$. This baby universe represents 5--D
FRW-like universe filled with dust and a negative cosmological
constant. It has a maximum at time corresponding to the wormhole
throat and bounces from a minimum occurring when the scale factor
equals to zero (usual FRW-type singularity). However, there is a
possibility for the analytical continuation to the Lorentzian
space-time which preserves branes. As it was pointed out in paper
\cite{Gariga-Sasaki}, it can be done for the case when 4--D
positive curvature metric $g^{(4)}$ in Eq.\rf{2.11} is the metric
of a 4--sphere:
\be{5.1} ds^2_E = dr^2 + a^2(r)(d\chi^2 + \sin^2\chi\,
d\Omega^2_{(3)})\, .\ee
Then, the analytic continuation of the azimuthal coordinate
$\chi$:
\be{5.2} \chi \longrightarrow iHt + \frac{\pi}{2}\, , \ee
results in a Lorentzian metric with evolving branes:
\be{5.3} ds^2_L = dr^2 + H^2 a^2(r)( - dt^2 + \frac{1}{H^2}\cosh^2
Ht \; d\Omega^2_{(3)})\, . \ee
Here, parameter $H$ is chosen in such a way that $t$ describes the
proper time on the brane. For example, in the one-brane-case
\rf{3.1}:
\be{5.4} \left. H a(r)\right|_{r=L} = 1 \quad \Longrightarrow
\quad H = \frac{1}{a(L)} = \frac{\sqrt{2}}{l} \left( \sqrt{b}\,
\cosh \frac{2L}{l} -1 \right)^{-1/2}\, .\ee
The compact version of solution \rf{3.1} describes a 4--D
spherical brane of radius $a(L) = H^{-1} $ enclosing 5--D space
with frozen dust ( equivalently, frozen pressureless scalar field)
and negative cosmological constant (AdS bulk). The south pole of
the 4--D sphere with the azimuth coordinate $\chi = 0$ can be
treated as "nothing". The birth of the brane-world takes place at
the time $t=0$ which corresponds in the Euclidean region to $\chi
= \pi /2$ (the equator of the 4--D sphere) where the sections
$\chi = \const$ of the 4--D sphere reach their maximum with radius
$H^{-1}$.
After birth, the brane represents an evolving 3--D sphere with
initial radius $H^{-1}$. More precisely, it is an inflating de
Sitter space with the Hubble constant $H$ (see figure 2).

In the case of $n$ branes described by expression \rf{3.3}, we can
introduce for each of the coordinate patches between the throats
$r_{(th)i} \le r \le r_{(th)i+1}$ the following transformation:
\be{5.5} \chi \longrightarrow iH_it_i + \frac{\pi}{2}\, , \ee
where
\be{5.6} H_i = \frac{1}{a(r_i)} \ee
and $r_i \in [r_{(th)i}, r_{(th)i+1}]$ is the position of the i-th
brane ($i=1,\ldots ,n$). After this continuation, the Lorentzian
metric is:
\be{5.7} ds^2_L = dr^2 + H_i^2 a^2(r)( - dt_i^2 +
\frac{1}{H_i^2}\cosh^2 H_it_i \; d\Omega^2_{(3)})\, ,\ee
where coordinates $t_i$ are glued to each other at the throats:
\be{5.8} t_i = \frac{a_i}{a_j}t_j\, . \ee
Thus, each of $n$ branes at $r=r_i$ represents a de Sitter space
with its own proper time $t_i$ and own Hubble constant $H_i$.

To conclude this section, we would like shortly comment some
alternative possibilities of the analytic continuation. As we have
noted before, the analytical continuation $r\to i\tau$ leads to
the creation of a 5--D baby universe with scale factor,
\be{5.9} a(\tau)= \frac{l}{\sqrt{2}}\left(\sqrt{b}
\cos\frac{2\tau}{l} - 1\right)^{1/2}\; , \quad 0 \le \tau \le
\frac{l}{2} \arccos \frac{1}{\sqrt{b}}\, . \ee
{}From this model we can actually construct a spherical 3--D brane
by using a similar procedure to that we have described in section
3; that is excising regions with $\chi > L < \pi $ for two
identical 4-sphere and gluing the remaining two copies along
3--spheres $\chi = L$. The brane-world model in this case is
described by the metric:
\be{5.10} ds^2_L = -d\tau^2 + a^2(\tau)( d\chi^2 + a^2(\chi )\;
d\Omega^2_{(3)})\, ,\ee
where the scale factor $a(\tau )$ is defined by Eq.\rf{5.9} and
the scale factor $a(\chi )$ reads
\be{5.11} a (\chi ) = \left\{\begin{array}{rcl} \sin \chi \quad
\quad \quad &\, , &\quad 0 \le \chi \le L < \pi
\\ \sin (2L-\chi) &\, , & \quad L \le \chi \le 2L
\\
\end{array}\right. \ee
In this model, both the additional space as well as 3--D brane
have the same dynamical behaviour which is described by
Eq.\rf{5.9}.

We turn finally to another interesting possibility consisting in
the simultaneous analytic continuation in two directions: $\chi
\to it + \pi/2$ and $r \to i\tau$. It results in a metric:
\be{5.12} ds^2_L = -d\tau^2 + a^2(\tau)(-dt^2 + \cosh^2t\;
d\Omega^2_{(3)})\, ,\ee
which describes a multidimensional/multitemporal solution of the
Einstein equations. Relative to time $\tau$ we have the FRW-type
universe \rf{5.9} and with respect to time $t$ we obtain the de
Sitter universe with the Hubble constant $H=1$. The number of
spatial coordinates here is equal to usual three. In the present
paper we shall not investigate in detail solutions \rf{5.10} and
\rf{5.12} postponing their study for future work.

\section{Volcano and multi-Volcano potentials\label{potential}}
\setcounter{equation}{0}

\bigskip

In this section we will discuss briefly the localization of
gravity at the branes described above. It is well known for
metrics of the form \rf{2.2} and \rf{2.11}, rewritten in the
conformal time gauge, that if tensor metric perturbations $\delta
g_{MN} (X) = h_{MN} (\eta ,x)$ is taken in the gauge: $h_{\eta
\eta} = h_{\eta \mu} = h_{\mu}^{\mu} ={h^{\mu \nu}}_{;\nu} = 0$
(where the semicolon denotes a covariant derivative with respect
to d--dimensional metric $g_{\mu \nu}^{(d)}(x)$) and is normalized
and factorized as $h_{\mu \nu} (\eta , x) = a^{-(d-5)/2}(\eta )
\psi (\eta )\hat h_{\mu \nu}(x)$, then the radial profile of the
perturbations satisfies a Schr$\ddot{o}$dinger-like equation
\cite{RS2}
\be{4.1} \left( -\frac{d^2}{d\eta^2} + V(\eta ) \right) \psi (\eta
) = M^2 \psi (\eta )\, , \ee
in which $V(\eta)$ is the so-called "volcano" potential
\be{4.2} V(\eta ) =\left. \left( \frac{d^2}{d\eta^2}
a^{\frac{d-1}{2}}\right) \right/ a^{\frac{d-1}{2}}\, , \ee
and $M$ defines a mass spectrum of spin--2 perturbations $\hat
h_{\mu \nu}(x)$ moving in the d--dimensional background metric
$g^{(d)}_{\mu \nu}(x)$. If the spectrum of mass starts from $M=0$,
then a state corresponding to this value of $M$ describes usual
4--dimensional (for $d = 4$) massless gravitons responsible for
the standard Newtonian gravity.

It can be seen that for $M=0$ the trivial solution of Eq.\rf{4.1}
is
\be{4.3} \psi \propto a^{\frac{d-1}{2}} \, .\ee
Thus, if the scale factor $a$ has maxima at branes, this zero-mode
state is localized at the branes. On the other hand, if the shape
of the volcano potential is such that it suppresses the
Kaluza--Klein modes with $M>0$ at the branes by the wings of the
potential, then the massless graviton gives the main contribution
to gravity on the brane and effectively we have the Newtonian
gravitation here. Moreover, to have a well defined effective
gravitation on branes, it is desirable to have a mass spectrum
where the massless graviton is well separated by a mass gap from
other massive Kaluza--Klein modes \cite{Nam}. This is precisely
the general situation that we have in the present paper for
compact models. Indeed, in the case of the compact n-brane model
described by \rf{3.3} the radial profile reads:
\be{4.7} \psi \propto a^{3/2}(r)= \sum_{i=1}^{n+1} a_i^{3/2} (r)\,
\theta_i (r)\, ,\ee
where we used the properties of the $\theta$--function (see
footnote \ref{5}). This formula shows that the massless graviton
(zero-mode) is localized on each of the branes as they approaches
their local maximum at the branes.

The volcano potential \rf{4.2} for this model can be written as
follows
\ba{4.8} V &=& a^{-3/2} \frac{d^2}{d\eta^2} a^{3/2} = a^{-3/2}
\left( a \frac{d a}{d r} \frac{d }{d r} a^{3/2} + a^2 \frac{d^2}{d
r^2} a^{3/2}\right)\nn \\ &=& \sum_{i=1}^{n+1} \overline
V_i(r)\theta_i (r)- \frac{16}{3} \kappa^2_5 \sum_{i=1}^{n} a_i^2 T
(r) \delta (r - r_i)\, , \ea
where
\be{4.9} \overline V_i(r) = \frac{15}{4}\sqrt{b} \cosh^2
(r-2\sum_{k=1}^{i-1}L_k) / l\; \; \frac{\sinh^2
(r-2\sum_{k=1}^{i-1}L_k)/l +(4/5)m_1}{\sinh^2
(r-2\sum_{k=1}^{i-1}L_k)/l + m_1} \, -\, \frac{3}{2}\sqrt{b} \; .
\ee
Here, we used the expressions \rf{3.3} and \rf{3.4} for the
functions $a_i$ and $\theta_i$ and the tension $T(r_i)$ has the
form of Eq.\rf{3.2} with the replacement $L \to L_i\, , \; i =
1,\ldots ,n$. It can be seen that, at the throats, the potential
\rf{4.8} reaches its minima $\left. V(r) \right|_{r_{(th)i}} \, ,
\; i= 1,\ldots ,n\, $:
\be{4.5} \left. V(r)\right|_{r_{(th)i}}\; \equiv \, V_{th}\, =\,
\left\{\begin{array}{rcl} (3/2) \sqrt{b} &\, , &\quad b \ne 1\\
9/4\quad &\, , & \quad b = 1
\\
\end{array}\right. \ee
and, at the branes, achieves its local maxima:
\be{4.10} V_{(br)i} = \frac{15}{4}\sqrt{b} \cosh^2 (L_i/l)
\frac{\sinh^2 (L_i/l) +(4/5)m_1}{\sinh^2 (L_i/l) + m_1} \, - \,
\frac{3}{2}\sqrt{b}\; , \quad i=1,\ldots ,n \, , \ee
which tend to $ (15/4) \cosh^2 (L_i/l) - 3/2$ in the limit $b\to
1$, corresponding to the vacuum case \rf{2.13}. Two different
values of the minimum expressed in Eq.~(\ref{4.5}) result from the
two different asymptotic behaviours of the function $\cd\,(u | m)
\to 0 /\sqrt{1-m}$ when $ u \to K(m)$. Hence, this fraction equals
to 0 if $m \ne 1$ and goes to $1$ when $m \to 1$. The second value
of Eq.~(\ref{4.5}) $V_1 = 9/4$ was obtained in
\cite{Gariga-Sasaki} for model \rf{2.13}. Thus, Kaluza--Klein
modes starts at
\be{4.6} M = \sqrt{V_{th}}\, , \ee
and this mass gap separates the zero-mode from other massive
Kaluza--Klein modes\footnote{It is clear, that to an observer on a
i-th brane at $r=r_i$, the physical metric is the induced metric
on this brane: $g^{(4)}_{(ph)\mu \nu} = a_i^2(r_i ) g^{(4)}_{\mu
\nu}$. It results in an appropriate rescaling of effective
physical values of, for example, effective 4--D cosmological and
gravitation constants on the brane (see \cite{Bouhmadi-Zhuk}).
Obviously, physical Kaluza--Klein masses for this observer are
similarly rescaled, $M \to m_{(ph)} = M/a_i(r_i )$. \label{7}}. It
is obvious that the spectrum of the Kaluza--Klein modes is
discrete for the compact models.

Thus, in the n-brane case, potential \rf{4.8} has the form of a
string (closed string for compact models) of volcano potentials
(see figure 3) and can be named as "multi-Volcano potential".
Obviously, this form of the potential entails suppression of the
massive Kaluza-Klein modes at the branes.

To conclude this section, we would like to remark an interesting
possibility arising from the special form of the multi-Volcano
potential with periodic structure. In this case we obtain a band
structure for the mass spectrum of the Kaluza-Klein modes and only
masses from these bands are allowed to exist (see e.g.
\cite{Nam}). This would give rise to separation between the
zero-mode and the massive Kaluza--Klein modes even larger than
those induced by the mass gap \rf{4.6}.


\section{Conclusions}

\setcounter{equation}{0}

\bigskip

In this paper we have considered in some detail 5--dimensional
cosmological models corresponding to the case of a bulk filled
with a pressureless scalar field, or equivalently by dust, and a
cosmological constant. Special attention has been given to the
case of a negative cosmological constant in a bulk with the
spatial geometry of a four-sphere. For this model, we have found
an asymptotically AdS wormhole instantonic solution. Brane-world
instantons with a single 4--D spherical brane as well as with a
string of such concentric branes can then be built up by using a
cutting and gluing procedure. We have been able to obtain regular
solutions which are free from any singularities at the origin of
extra coordinates, and can be compactified so that the asymptotic
divergences of the scale factor are prevented. Zero-mode massless
gravitons are shown to be localized on these 4--D branes, so
allowing such branes to nest Newtonian gravity. Analytical
continuation from the brane instantonic metric to the Lorentzian
regime leads to de Sitter 4--dimensional inflating branes. After
birth of the inflating brane world from "nothing", the perfect
fluid (dust) remains frozen: it is contained in the bulk but not
on the brane. Here, inflation has pure geometrical origin: the
Hubble constant of the inflating brane world is defined by the
inverse radius of the Euclidean 4--D spherical brane.

Other brane-world models with similar proprieties, as the ones
discussed in the sections \ref{instanton} and \ref{birth}, can be
obtained following the procedure explained in the present paper.
For example, inflating branes can be constructed when the 5--D
space has positive or zero cosmological constant, whenever the
topology of the instanton is $\mathbb{R}\times S^{4}$. Moreover,
using these instantons, we can get a class of brane-world models
characterized by a common dynamical behaviour for the
extra-dimension and brane. On the other hand, brane-world models
with static flat brane or with a number of such parallel branes
can be described considering an instanton with negative
cosmological constant and flat 4--D spatial topology ($k=0$).

Additionally, our model can be generalized for an arbitrary number
of dimensions. This can be performed by considering the model
(\ref{2.2}) with topology $\mathbb{R} \times M^{d}$. Here, the
d-dimensional manifold $M^{d}$ undergoes a topological splitting
into $n$ Einstein spaces: $M^{d} = \prod_{i=1}^n M_i^{d_i}\, , \;
g^{(d)} = \sum_{i=1}^n g^{(d_i)}_{(i)}\, ,\; R_{\mu
\nu}[g^{(d_i)}_{(i)}] = \lambda_i g^{(d_i)}_{(i)\mu \nu}\, ,\; d =
\sum_{i=1}^n d_i$, with $\lambda_i > 0$. Now, using the conformal
transformation $g^{(d_i)}_{(i)} \to
(\lambda_i/\lambda)g^{(d_i)}_{(i)}$, it can be easily shown
\cite{topsplit} that the constituent manifold $M^d$ is also the
Einstein space: $R_{\mu \nu}[g^{(d)}] = \lambda g^{(d)}_{\mu \nu}$
and the scale factor $a(\tau)$ satisfies Eq. (\ref{2.9}). Thus,
the brane-world instantons, as well as the birth of the
brane-worlds from them, can be constructed for this
$(d+1)$-dimensional model in complete analogy to the sections
\ref{instanton} and \ref{birth}. For example, in the particular
case $M^d = S^4 \times \prod_{i=1}^n S^{d_i}$, the analytic
continuation (\ref{5.2}) with respect to the coordinate $\chi$ of
4--sphere $S^4$ results in the following Lorentzian metric:
\be{c1}ds^2_L = dr^2 + H^2 a^2(r)( - dt^2 + \frac{1}{H^2}\cosh^2
Ht \; d\Omega^2_{(3)} + \frac{r_1^2}{H^2}\; d\Omega^2_{(d_1)} +
\ldots + \frac{r_n^2}{H^2}\; d\Omega^2_{(d_n)})\, ,\ee
where $r_i\, , \; i= 1, \ldots ,n\; $ are the radii of
$d_i$-spheres. We arrive at the brane-world model with inflating
4--D part of the brane, plus frozen, compactified and unobservable
(for $r_i \lesssim 10^{-17}cm$) $(d-4)$ dimensions on it. This
scenario, with a brane of codimension one, is of interest
because:\\ 1. it demonstrates the very interesting possibility for
the unification of new brane-world scenarios with the standard
Kaluza--Klein approach.\\ 2. it gives a possibility for the
localization of the gauge fields on the brane \cite{DRT},
\cite{Rubakov}.

Of course, some important aspects of this research need further
investigations including the study of the transition from
inflationary branes into a matter-dominated brane universe, the
physical consequences from the above mentioned brane-world model
with dynamical equivalence between extra dimension and branes, as
well as the meaning of the 5-dimensional space-times described by
a metric with two timelike dimensions.


\bigskip
{\bf Acknowledgments}

A.Z. thanks Instituto de Matem\'{a}ticas y F\'{\i}sica Fundamental, CSIC,
for kind hospitality during preparation of this paper. A.Z.
acknowledges support by Spanish Ministry of Education, Culture and
Sport (the programme for Sabbatical Stay in Spain) and the
programme SCOPES (Scientific co-operation between Eastern Europe
and Switzerland) of the Swiss National Science Foundation, project
No. 7SUPJ062239. M.B.L. is supported by a grant of the Spanish
Ministry of Science and Technology. This investigation was
supported by the GICYT under Research Project No. PB97-1218.




\newpage

\noindent
\begin{picture}(10,8.4)
\epsfxsize=13cm
\epsfbox{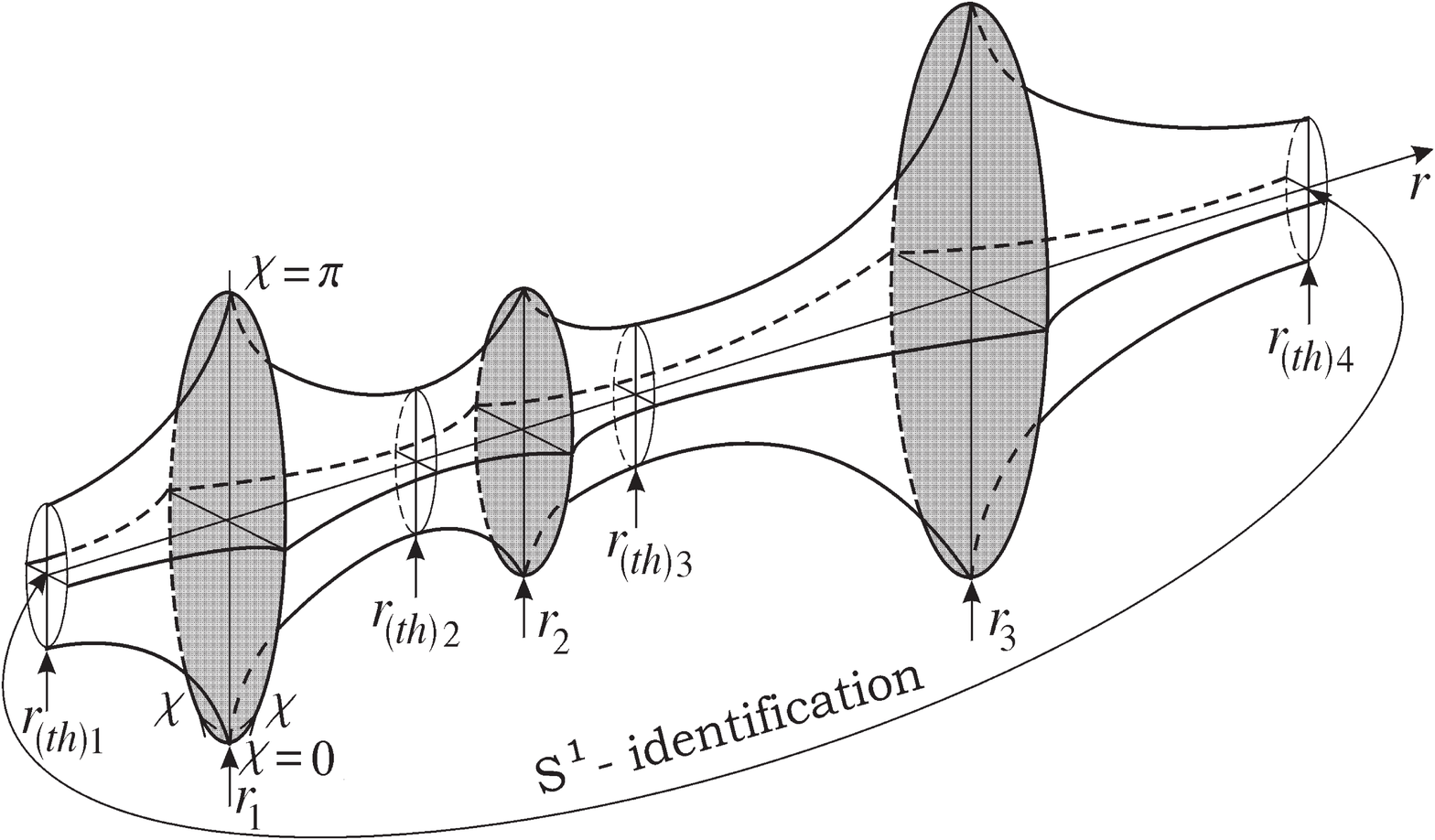}
\end{picture}

Figure 1: Compact n-brane-world instanton \rf{3.3} in the
case $n=3$. Each point in the figure represents a 3--D sphere. Lines with
fixed coordinates $r = \const \equiv r_0$ correspond to 4--D spheres with
radii $a(r_0)$. The branes are 4--D concentric spheres surrounding the 5--D
bulk.

\vspace*{1cm}

\noindent
\begin{picture}(10,9.5)
\epsfxsize=13cm \epsfbox{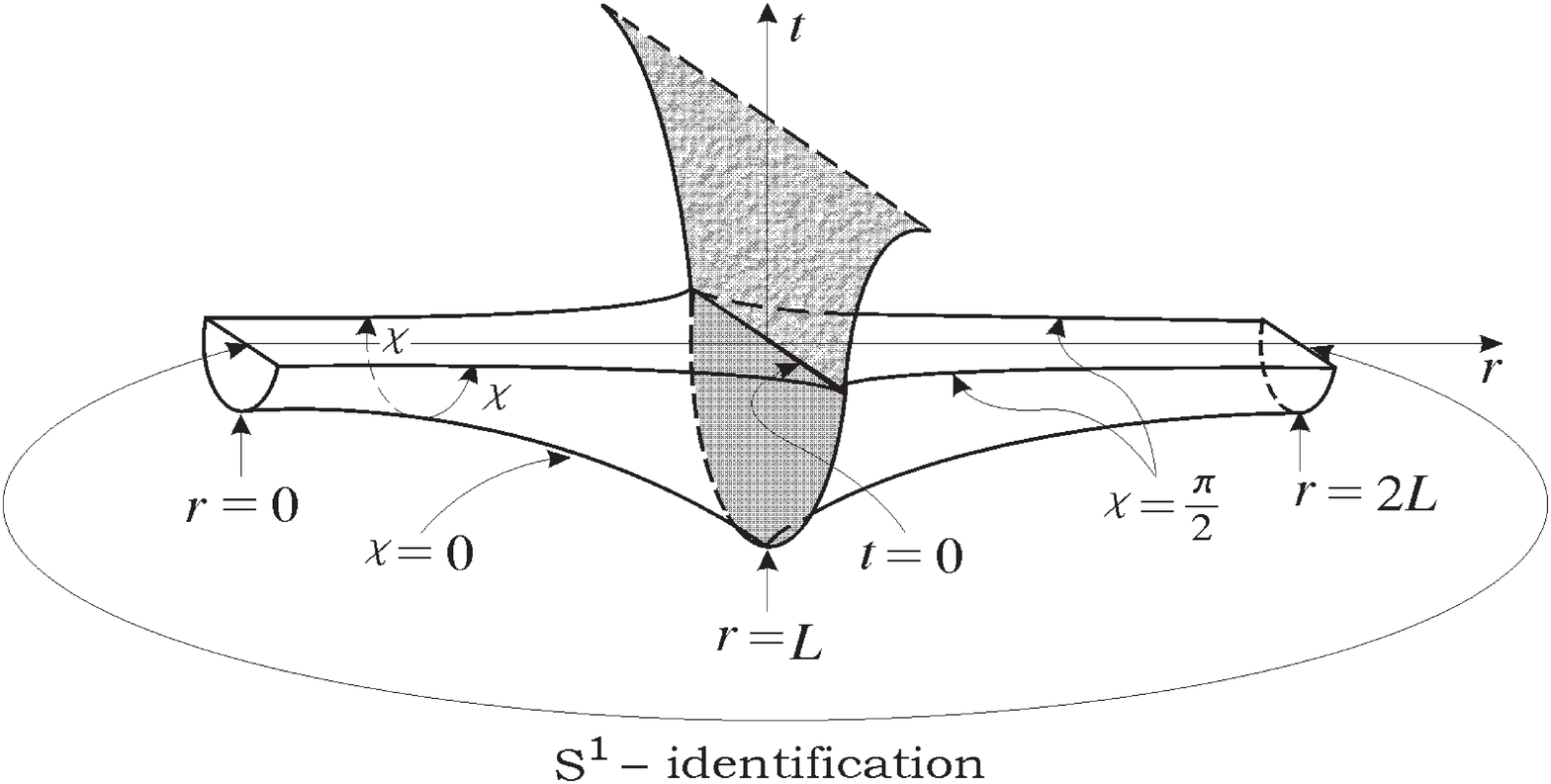}
\end{picture}

Figure 2: Birth of the one-brane-world $(n=1)$ from ``nothing''.
The creation takes place at the time $t=0$ and the 5--D Lorentzian
metric is described by Eq.\rf{5.3}. The brane after birth
represents an inflating 3--D sphere with initial radius $H^{-1}$
that is de Sitter space-time with Hubble constant $H$.

\newpage

\vspace*{2cm}

\noindent
\begin{picture}(10,9.5)
\epsfxsize=11cm
\epsfbox{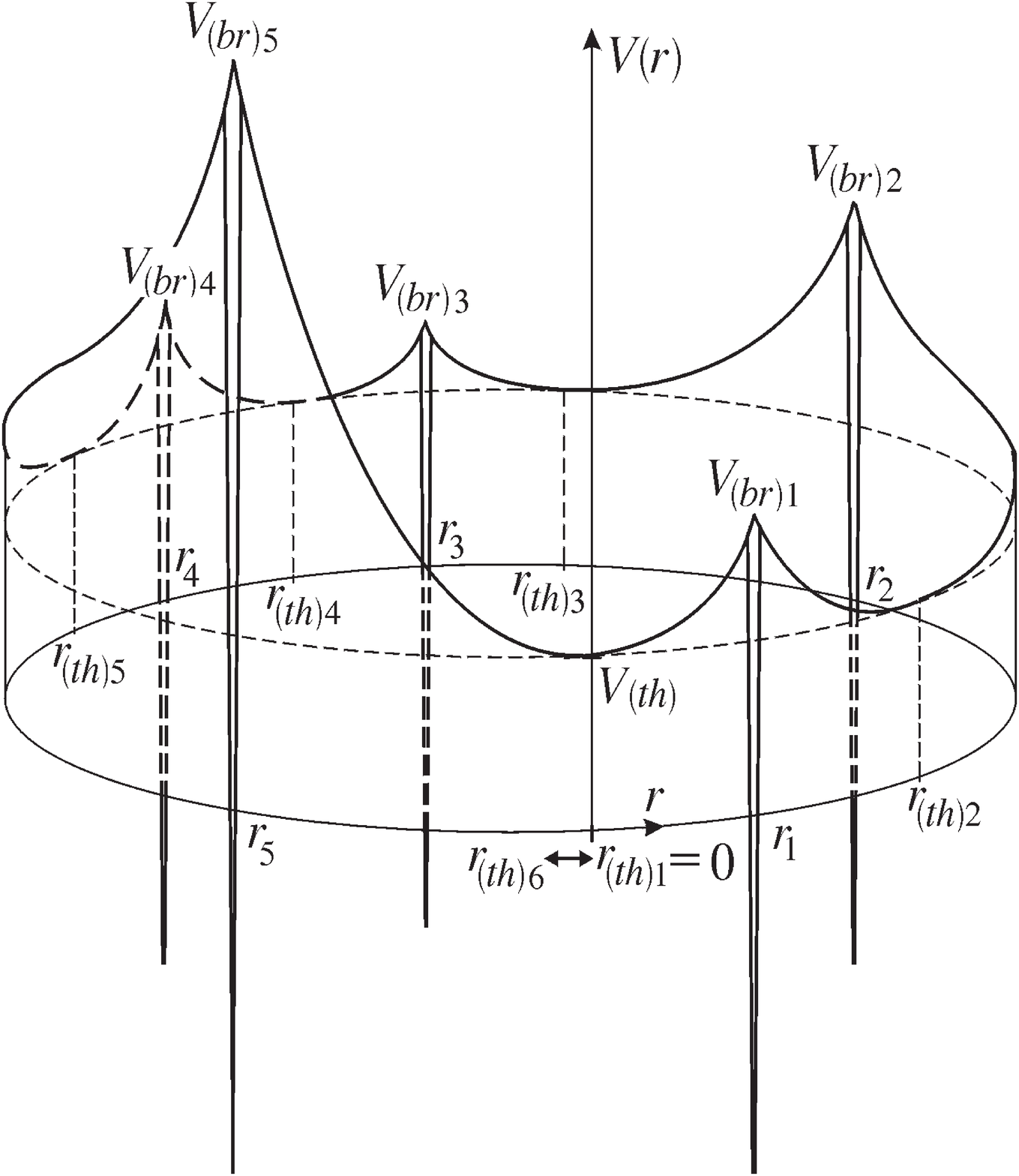}
\end{picture}

Figure 3: Multi-Volcano potential \rf{4.8} for the compact
5--brane-world instanton when $n=5$. Each of the "volcanoes" has a
local maximum $V_{(br)i}$ localized on the branes. Identical local
minima $V_{th}$ correspond to the wormhole throat positions
$r_{(th)i}$.

\end{document}